\def\useextern{}

\documentclass[runningheads]{llncs}
\usepackage[T1]{fontenc}
\usepackage{graphicx}
\usepackage{comment}
\usepackage{tabularx}
\usepackage[nameinlink,noabbrev,capitalize]{cleveref}
\usepackage{caption}
\usepackage{subcaption}
\usepackage{booktabs}
\usepackage{xspace}
\usepackage{xcolor}

\usepackage{ifluatex}
\ifluatex
\usepackage[newfloat,finalizecache]{minted}
\else
\usepackage[newfloat,frozencache]{minted}
\fi
\newminted{c}{fontsize=\small,escapeinside=``}
\newminted{cpp}{fontsize=\small,escapeinside=``}
\newminted{text}{fontsize=\small,escapeinside=``}
\newmintinline{c}{style=bw}
\newmintinline{cpp}{style=bw}
\newmintinline{text}{style=bw}

\usepackage{pgfmath}

\ifdefined\useextern
\usepackage{tikzexternal}
\else
\ifdefined\genextern
\usepackage{tikz}
\usetikzlibrary{external}
\else
\usepackage{tikz}
\fi

\usetikzlibrary{calc}
\usetikzlibrary{decorations}
\usetikzlibrary{decorations.pathreplacing}
\usetikzlibrary{decorations.text}
\usetikzlibrary{decorations.pathmorphing}
\usetikzlibrary{decorations.markings}
\usetikzlibrary{fit}
\usetikzlibrary{shadows.blur}
\usetikzlibrary{arrows}    
\usetikzlibrary{shapes}
\usetikzlibrary{shadows}
\usetikzlibrary{automata}
\usetikzlibrary{positioning}
\usetikzlibrary{petri}
\usetikzlibrary{chains}
\usetikzlibrary{fadings}
\usetikzlibrary{matrix}
\usetikzlibrary{patterns}
\usetikzlibrary{mindmap}
\usetikzlibrary{graphs}			
\usetikzlibrary{graphdrawing}	
\usetikzlibrary{quotes}			
\usetikzlibrary{babel}          
\usetikzlibrary{hobby}
\usetikzlibrary{arrows.meta}
\usetikzlibrary{backgrounds}
\usetikzlibrary{bending}

\usegdlibrary{trees}			
\usegdlibrary{layered}			
\usegdlibrary{force}			
\usegdlibrary{circular}			
\usegdlibrary{routing}			

\pgfdeclarelayer{backbackbackground}
\pgfdeclarelayer{backbackground}
\pgfdeclarelayer{background}
\pgfdeclarelayer{foreground}
\pgfdeclarelayer{foreforeground}
\pgfsetlayers{backbackbackground,backbackground,background,main,foreground,foreforeground}
\tikzset{>={Stealth[round,flex,length=5pt 4.5 0.8]}} 
\tikzset{tight/.style={minimum width=0pt,minimum height=0pt,inner sep=0pt,outer sep=0pt}}

\fi

\newcommand*{\dquote}[1]{``#1''}
\newcommand*{\squote}[1]{`#1'}

\newcommand*{\textreset}[1]{{\normalsize\rm\textcolor{black}{#1}}}

\newcolumntype{P}{X}								
\newcolumntype{L}{>{\raggedright\arraybackslash}X}	
\newcolumntype{C}{>{\centering\arraybackslash}X}	
\newcolumntype{R}{>{\raggedleft\arraybackslash}X}	

\definecolor{babyblueeyes}{rgb}{0.63, 0.79, 0.95}
\definecolor{jonquil}{rgb}{0.98, 0.85, 0.37}

\makeatletter

\newsavebox{\my@scale@Lrbox}
\newcommand*{\my@scale@Percentage}{}
\newenvironment*{scaleenv}[1]{%
\renewcommand*{\my@scale@Percentage}{#1}%
\begin{lrbox}{\my@scale@Lrbox}%
}{%
\end{lrbox}%
\scalebox{\my@scale@Percentage}{\usebox{\my@scale@Lrbox}}%
}%

\newsavebox{\my@scalepar@TempBox}
\newenvironment{scalepar}[1]{%
\def\my@DoScalePar{\scalebox{#1}}%
\begin{lrbox}{\my@scalepar@TempBox}%
\pgfmathparse{\textwidth/#1}%
\begin{minipage}{\pgfmathresult pt}%
}{%
\end{minipage}%
\end{lrbox}%
\my@DoScalePar{\usebox{\my@scalepar@TempBox}}%
}

\newsavebox{\my@resizeenv@TempBox}%
\newcommand*{\my@resizeenv@width}{}%
\newenvironment{resizeenv}[1]{%
\renewcommand*{\my@resizeenv@width}{#1}%
\begin{lrbox}{\my@resizeenv@TempBox}%
}{%
\end{lrbox}%
\resizebox{\my@resizeenv@width}{!}{\usebox{\my@resizeenv@TempBox}}%
}%

\newenvironment{resizepar}{%
\begin{resizeenv}{\textwidth}%
}{%
\end{resizeenv}%
}%

\makeatother

\makeatletter
\tikzset{
    database top segment style/.style={draw},
    database middle segment style/.style={draw},
    database bottom segment style/.style={draw},
    database/.style={
        path picture={
            \path [database bottom segment style]
                (-\db@r,-0.5*\db@sh) 
                -- ++(0,-1*\db@sh) 
                arc [start angle=180, end angle=360,
                    x radius=\db@r, y radius=\db@ar*\db@r]
                -- ++(0,1*\db@sh)
                arc [start angle=360, end angle=180,
                    x radius=\db@r, y radius=\db@ar*\db@r];
            \path [database middle segment style]
                (-\db@r,0.5*\db@sh) 
                -- ++(0,-1*\db@sh) 
                arc [start angle=180, end angle=360,
                    x radius=\db@r, y radius=\db@ar*\db@r]
                -- ++(0,1*\db@sh)
                arc [start angle=360, end angle=180,
                    x radius=\db@r, y radius=\db@ar*\db@r];
            \path [database top segment style]
                (-\db@r,1.5*\db@sh) 
                -- ++(0,-1*\db@sh) 
                arc [start angle=180, end angle=360,
                    x radius=\db@r, y radius=\db@ar*\db@r]
                -- ++(0,1*\db@sh)
                arc [start angle=360, end angle=180,
                    x radius=\db@r, y radius=\db@ar*\db@r];
            \path [database top segment style]
                (0, 1.5*\db@sh) circle [x radius=\db@r, y radius=\db@ar*\db@r];
        },
        minimum width=2*\db@r + \pgflinewidth,
        minimum height=3*\db@sh + 2*\db@ar*\db@r + \pgflinewidth,
    },
    database segment height/.store in=\db@sh,
    database radius/.store in=\db@r,
    database aspect ratio/.store in=\db@ar,
    database segment height=0.1cm,
    database radius=0.25cm,
    database aspect ratio=0.35,
    database top segment/.style={
        database top segment style/.append style={#1}},
    database middle segment/.style={
        database middle segment style/.append style={#1}},
    database bottom segment/.style={
        database bottom segment style/.append style={#1}}
}
\makeatother

\begin{document}
\title{Generalizing Hierarchical Parallelism}
%
%
\author{Michael Kruse\orcidID{0000-0001-7756-7126}}
\authorrunning{M. Kruse}
%
\institute{Argonne National Laboratory\\Mathematics and Computer Science Division\\9700 S. Cass Avenue, Lemont, IL 60439, USA\\
\email{michael.kruse@anl.gov}}
\maketitle              
\begin{abstract}
Since the days of OpenMP 1.0 computer hardware has become more complex, typically by specializing compute units for coarse- and fine-grained parallelism in incrementally deeper hierarchies of parallelism.
Newer versions of OpenMP reacted by introducing new mechanisms for querying or controlling its individual levels, each time adding another concept such as places, teams, and progress groups. 
In this paper we propose going back to the roots of OpenMP in the form of nested parallelism for a simpler model and more flexible handling of arbitrary deep hardware hierarchies.

\end{abstract}


\section{Introduction}

Contemporary hardware architecture has changed significantly since OpenMP was introduced.
OpenMP 1.0 was designed for symmetric multiprocessing (SMP) systems, when processors could run at most one thread~\cite{openmp10}.
There was no hierarchy, as implied by \emph{symmetric}: Everything was at the same level without differences in performance or communication between any two CPUs.

The programming model therefore was comparatively simple: one directive to start (thread-)parallelism (\cinline{#pragma omp parallel}) and execute its associated region as a single program multiple data (SPMD) instance, two directives to distribute work between them (\cinline{#pragma omp for} and \cinline{#pragma omp sections}), and some directives such as barriers and memory fences.
It was intended mainly to standardize proprietary compiler extensions that various companies introduced.
It was possible to establish a logical hierarchy by executing a parallel directive inside another parallel-construct, but this has no performance advantage over using all threads in a single parallel construct and makes it difficult to not under- or oversubscribe the available hardware.

\begin{table}[t]
\begin{scalepar}{0.9} 
\begin{tabularx}{\textwidth}{cCc}
    Hardware Feature                                                                           & OpenMP Feature                                        & Version    \\
    \cmidrule(r){1-1} \cmidrule(lr){2-2} \cmidrule(l){3-3}
    Symmetric Multiprocessing (SMP)                                                            & \texttt{\#pragma omp parallel for}                    & 1.0        \\
    scheduling with dependencies                                                               & \texttt{\#pragma omp task}                            & 3.0        \\
    Non-Uniform Memory Access (NUMA)                                                           & \texttt{OMP\_PROC\_BIND}, \texttt{OMP\_PLACES}        & 3.1, 4.0   \\
    Cores, shared caches                                                                       & \texttt{OMP\_PROC\_BIND}, \texttt{OMP\_PLACES}        & 3.1, 4.0   \\
    Symmetric Multithreading (SMT)                                                             & \texttt{OMP\_PROC\_BIND}, \texttt{OMP\_PLACES}        & 3.1, 4.0   \\
    Single Instruction Multiple Data (SIMD)                                                    & \texttt{\#pragma omp simd}                            & 4.0        \\
    Heterogeneous Accelerators                                                                 & \texttt{\#pragma omp target}                          & 4.0        \\
    GPGPU Multiprocessors                                                                      & \texttt{\#pragma omp teams distribute}                & 4.0        \\
    Single Instruction Multiple Threads (SIMT)                                                 & \texttt{\#pragma omp parallel for}, \texttt{safesync} & 1.0, 6.0 
\end{tabularx}
\end{scalepar}%
\bigskip%
\caption{Changes in compute architectures and how OpenMP addressed them.}%
\label{tab:ompreaction}%
\vspace*{-7mm}%
\end{table}

This model was not sufficient anymore after the hardware became more complex and introduced NUMA, multiple cores per CPU, SMT, and SIMD.
While applications could just ignore these characteristics and continue to work correctly, these applications would not be able to reach the best possible performance.
Therefore, shown in \cref{tab:ompreaction}, OpenMP 3.1 introduced the \cinline{proc_bind} mechanism, affinity, and other features to account for work placement to hardware units.

The assumption that everything is an abstraction of SMP threads then was broken with the advent of GPGPUs.
Nvidia GPUs split the execution into grid, block, and warp and introduced the CUDA programming language, which abstracted hardware blocks and warps into the single instruction multiple threads (SIMT) model, but the grid remained separate because of not supporting fundamental mechanisms across processing units such as barriers.
OpenMP could no longer be used as a programming model; only Intel's Xeon Phi many-core accelerator held up the global cache coherency required by OpenMP.
Consequently, OpenMP added two new layers: \cinline{target} for device memory and \cinline{teams} for computations with fewer synchronization guarantees than threads to match the CUDA model.
Unlike traditional threads, SIMT threads are executed in lockstep which was a breaking change since older OpenMP applications may have assumed a fair scheduler.
The current draft of OpenMP 6.0 takes this into accounting with a \cinline{safesync} clause.

The current hardware generation has many more levels that we explore in \cref{sec:hardware}.
For instance, most vendors these days combine multiple chiplets into a single packaging, where communication between chiplets is necessarily slower than within a single die, i.e. NUMA but for \cinline{teams}~\cite{geist22-frontier}.
We expect that future processors will be even more complex and its consequences in terms of performance even more noticeable.

Also, even today most high-performance computing platforms have multiple GPUs per node; but since  OpenMP has no constructs to represent this level of parallelism, each GPU has to be targeted separately.
One proposal is to another pair of constructs: \cinline{leagues} to start multiple sets of teams (one set per GPU) and \cinline{spread} for work distribution between them.
The composite directive for embarrassingly parallel code using all available parallelism therefore would become\footnote{Our proposal is \texttt{\#pragma omp parallel for level(devices,teams,threads,lanes)}}
\begin{minted}[fontsize=\small,escapeinside=||]{c}
#pragma omp target leagues spread teams distribute parallel for simd |\textreset{.}|
\end{minted}
Another proposal is to make a league span over teams from different GPUs, either by the compiler treating them as such, or the runtime offering a \squote{superdevice} combining multiple GPUs. 
In addition to creating another NUMA problem when \textinline{OMP_PLACES} currently applies only to host code, GPUs may not support all the synchronization mechanisms across GPUs, such as atomics.
This may actually change in the future, similar to most GPUs now support these mechanisms on the \cinline{teams}-level.

Since the term \emph{thread} has many meanings that depend on context, in the following we avoid this term.
We will use \emph{task} to mean an execution of an SPMD region\footnote{The meaning in OpenMP is \dquote{instance of executable code and its data environment}, which also includes non-SPMD regions such as in the \texttt{task}-construct}, \emph{warp} to mean a collection of tasks that may execute in lockstep (even on non-Nvidia hardware), and \emph{lane} to denote one of the tasks executed in a warp.

\section{Algorithms for Using Multiple Levels}

Beyond individual algorithms that are specifically optimized to make use of a specific level's feature --- such as lookup-tables that fits a specific level's local memory --- some classes of algorithms can be implemented recursively and profit from any number of levels.
One of the most elementary is an implementation of a reduction: on each level, one of the tasks collects the results from all sibling tasks.
The most efficient means of communication on each level is chosen, for example, shuffle instructions on the warp level, to reduce the amount of work on the slower parent levels.
OpenMP has a reduction clause, but for user-defined operators, it does not permit non-commutativeness or making use of level-specific optimizations such as CUDA's \cinline{__shfl_xor_sync}.

Another class of algorithms is stencils.
In addition to profiting from warp-level instructions~\cite{wang20-ompshuffle}, they can be tiled to an arbitrary level~\cite{iwomp18}.
A region of stencil computations can be chosen such that their input region just fits into the local memory. 
This can be repeated on any level that has local memory.

Other algorithms making use of multiple levels are tensor comprehensions including matrix-matrix multiplications~\cite{low16} and butterfly-access patterns such as exposed by fast Fourier transforms~\cite{franchetti09-ol}.
In both cases, optimized algorithms select a subset of computations with data reuse and for each level ensure that this data is in local memory.

\section{Language Extensions for Hierarchical Parallelism}

The idea we propose is to reuse the mechanism of nested parallelism from OpenMP 1.0 but allow the nested construct to choose where to get the new parallelism from.
In this section we explore various aspects of controlling where a computation runs, and we continue in \cref{sec:mem} on using levels with local memory.

\subsection{Explicitly Selecting a Level}

The \emph{level} clause selects the level of parallelism to use, as shown below.
\begin{ccode}
#pragma omp parallel level(devices(0,1))
  #pragma omp parallel level(multiprocessors)
  {
    printf("Hello from multiprocessor %d\n", omp_get_thread_num());
    #pragma omp parallel level(warps)
    {
      #pragma omp parallel level(lanes)
      printf("Hello from lane %d of warp %d\n", omp_get_thread_num(), \
          omp_get_ancestor_thread_num(2));      
  } }
\end{ccode}
One of the motivations of this proposal is allowing performance optimization of specific devices, hence the arguments of the clause can be implementation-specific.
It is sensible to also specify a set of predefined levels for device-independent code that match the units of parallelism of the current OpenMP specification: \cinline{devices}, \cinline{teams}, \cinline{threads}, and \cinline{simd}.
The example above shows that the clause argument may take options; in the case of \cinline{devices} it selects which devices participate in the SPMD region.
These could also be filtered by a selector; for example, \cinline!device={isa("nvptx")}! would select all CUDA-based devices.
The \cinline{teams} and \cinline{threads} arguments correspond to the definition of the current teams and parallel directives, respectively;

\begin{table}[b!]
\begin{scalepar}{0.9} 
\begin{tabularx}{\textwidth}{p{9em}L}
\multicolumn{1}{c}{\textbf{Name}} &  \multicolumn{1}{c}{\textbf{Description}} \\
\cmidrule(r){1-1} \cmidrule(l){2-2}
\texttt{barrier}, \texttt{critical}, \texttt{atomic} & Whether this named synchronization directive is supported on this level.  \\
\texttt{shuffle} & Whether shuffle instructions are supported on this level; see \cref{sec:shuffle}.  \\
\texttt{oversubcribable} & Whether more tasks can be launched than the hardware can execute in parallel. In this case the operating system has to time-share the resource. \\
\texttt{dynamic} & Whether nonstatic schedules are allowed; see \cref{sec:scheduling}. \\
\texttt{lockstep}, \texttt{progress} & Whether the tasks execute in lockstep. If not, whether there is a forward progress guarantee. \\
\texttt{globalmem} & Whether the level has access to the global address space. If not, explicit mapping using the \cinline{map} clause is required; see \cref{sec:mem}. \\
\texttt{localmem} & Whether there is local memory on this level that can  be accessed efficiently but only by task in this level. \\
\texttt{groupmem} & Whether there is memory that has different contents for each sibling on this level, such as \cinline{threadprivate} and its generalization \cinline{groupprivate}. \\
\texttt{cache} &  Whether there is a transparent cache on this level. \\
\texttt{num(}\textit{c}\texttt{)} & Number of tasks this level can run in parallel. \\
\texttt{grainedness(}\textit{r}\texttt{)} & Measure of how long tasks on this levels should execute. The more nested the hardware level, the smaller the graininess should be, meaning fewer synchronizable tasks but also has less synchronization overhead. \\
\end{tabularx}
\end{scalepar}
\bigskip
\caption{Level properties.}%
\label{tab:sync}%
\vspace*{-7mm}%
\end{table}

Levels can be collapsed to form a single, encompassing level.
For instance,
\begin{ccode}
#pragma omp parallel level(devices(0-3),teams,threads)
\end{ccode}
executes the tasks using all threads in all teams of the selected devices.
Only synchronization guarantees that are common for all levels are also guaranteed for the collapsed levels.
Implementations may use aliases for common combinations of levels.
For instance, \cinline{level(teams)} could be considered an alias for \cinline{level(partitions,gpcs,tpcs,multiprocessors,tbps,ctas)} to hide the hardware-level details of Nvidia H100 GPUs.

In order to preserve compatibility with current OpenMP programs, the default level-clause argument has to be \cinline{threads}.
There is the potential for \cinline{#pragma omp parallel level(teams)} to subsume the semantics of \cinline{#pragma omp teams} including its combined and composite directives.

\subsection{Selecting Levels by Property}\label{sec:sync_clause}

The level clause (and teams/parallel constructs) assumes the programmer already has an idea about how the algorithm should execute, but this should not be relevant for first designing an algorithm.
Alternatively, the programmer could just define what features are needed for an algorithm, and then the compiler or runtime can select the appropriate hardware and levels to use.
The following code indicates that it needs a working implementation of a barrier.
\begin{ccode}
#pragma omp parallel sync(barrier)
{
  [...]
  #pragma omp barrier
  [...] 
}
\end{ccode}
OpenMP's teams directive does support a barrier that works across team boundaries, so only parallelism within a team could be used.
However, many targets support barriers on the hardware level that teams maps to and therefore use this level of parallelism.
\cinline{barrier} is an example of a level property, but many other properties could be defined; more possible properties are listed in \cref{tab:sync}.

The properties as defined here are \emph{positive} properties.
This contradicts the current semantics of the parallel directive, which, for example, has to support barrier directives even without \cinline{sync(barrier)}.
Hence, a \cinline{nobarrier} clause would be needed instead, to indicate the code does not need this feature.
However, we think it would be preferable if users would specify the features they need, rather than having to specify for each directive the list of features that are not used.
We propose to apply the default semantics of the parallel directive only if the sync clause is not present; that is, \cinline{sync()} has a different meaning from not specifying the clause at all.

\subsection{Reserving Nested Parallelism}

By default, \cinline{#pragma omp parallel sync()} would use all available parallelism, but with nested parallel constructs some of the hardware parallelism needs to be reserved for the inner levels.
This can be done with a reserve clause.
\begin{ccode}
#pragma omp parallel sync() reserve(sync(barrier))
  #pragma omp parallel sync(barrier)
  [...]
\end{ccode}
The outer directive uses all levels except the ones that match the reserve clause argument, but implementations should make at least some parallelism available to each level, if necessary by subdividing it (see \cref{sec:sublevels}).
OpenMP's current teams and parallel constructs match this example.
In addition to the sync argument, the level to be reserved could be selected explicitly by using the level-clause-style argument.
Multiple levels can be reserved by using multiple arguments to the reserve clause.

Implementing this may not be trivial, however, since at compile time it may not be known what kind of parallelism has been reserved for the inner construct. 
In the example above, the threads level and the simd level provide support barriers but are compiled  differently.
It may just compile the region for both of them, which is then selected by the runtime.
Some restrictions may be needed to limit the number of versions when selecting levels by property, especially for numeric variations such as by the simdlen clause.

\subsection{Work Distribution}

Currently, the worksharing loop and sections directives always bind to the innermost parallel construct, and the distribute directive always binds to the innermost teams construct.
With parallel gaining the functionality of both and more levels, it needs to gain the ability to bind to any surrounding parallel construct.
Our proposal suggests using a \cinline{bind_ancestor} clause, which is similar to the \cinline{bind} clause but takes an argument analogous to \mbox{\texttt{omp\_get\_ancestor\_thread\_num(}\textit{level}\texttt{)}}.
The shorter variant would be \cinline{bind(-1)}, which takes an argument relative to the current level.
\begin{ccode}
#pragma omp parallel sync() reserve(sync())
  #pragma omp parallel sync() 
    #pragma omp for bind_ancestor(0) // bind to outermost parallel
    for (int i = 0; i < 128; ++i)
      #pragma omp for // bind to innermost parallel
      for (int j = 0; j < 64; ++j)
        [...]
\end{ccode}

The reader may have noticed that the example first starts two levels of parallelism  and then work-distributes both levels independently.
This is contrary to how it currently needs to be written in OpenMP and to how OpenMP defines composite constructs.
\begin{ccode}
#pragma omp teams
  #pragma omp distribute
  for (int i = 0; i < 128; ++i)
    #pragma omp parallel
      #pragma omp for 
      for (int j = 0; j < 64; ++j)
        [...]
\end{ccode}
Written this way, the overhead of the parallel directive applies to each iteration of the outer loop.
The LLVM implementation tries to avoid this by compiling \cinline{#pragma omp teams distribute parallel for} into the former code where the associated loop is strip-mined by the number of teams, even though currently not expressible in OpenMP.
LLVM calls this form \squote{SPMD-mode}.

\subsection{Scheduling}\label{sec:scheduling}

The possibility of worksharing any level also allows adding a schedule clause on any level, including dynamic schedules.
OpenMP's distribute construct allows only static schedules since dynamic schedules would require expensive synchronization between teams.

Additionally, we propose \cinline{schedule(none)}, which is useful for shuffle instructions (\cref{sec:shuffle}).
In contrast to a static schedule, this ensures that there is only a single chunk.
Behavior is undefined if there are more logical iterations than tasks, to support the generation of efficient code.
Non-participating executions only have to be masked out.
If the loop is normalized (i.e., starting at 0 and incremented by 1), then the loop counter variable is identical to \cinline{omp_get_thread_num()}, and the runtime call (or introducing special variables such as \cinline{threadidx.x} in CUDA) can be avoided.

\subsection{Compiler-Transformation-Based Directives}\label{sec:simd_and_loop}

While the \cinline{teams distribute} and \cinline{parallel for} directives follow the \emph{start parallelism} and \emph{work-distribute} approach, the simd and loop directives do not.
These directives do both at the same time but at the cost of less programmer control.
Our proposed parallel directive can first initialize a vector context for a number of lanes, as shown here. 
\begin{ccode}
#pragma omp parallel level(simd(8)) private(partial_sum)
{
  #pragma omp for bind(simd) schedule(static,8)
  for (int i = 0; i < 32; ++i)
    partial_sum += A[i];
  [...]
}
\end{ccode}
The code first initializes all elements of the vector \cinline{partial_sum} with 0; then each lane sums up $\frac{32}{8}=4$ values from array \cinline{A}, a typical start of an implementation of a reduction. 
This resembles the SIMT execution model, there is no good reason why the SIMT programming model should be reserved for GPUs only.
For LLVM, the command line flag \textinline{-fopenmp-target-simd} was proposed that would make the simd construct map to the individual lanes of a warp~\cite{llvm21-ompsimd}.

Combining starting parallelism and work distribution can be done with the \cinline{parallel for} combined construct.
To get the descriptive semantics of the loop construct, one would simply use \cinline{#pragma omp parallel for sync()}.

\subsection{Warp-Level Primitives}\label{sec:shuffle}

Another feature that CUDA supports but OpenMP does not is shuffle instructions, even though support has been proposed~\cite{wang20-ompshuffle}.
One of the difficulties is that a parallel directive may not be mapped to a hardware layer that  supports them; and even if it does, warp sizes vary between devices.
The following shows a solution based on our proposal.
\begin{samepage}
\begin{ccode}
#pragma omp parallel sync(shuffle,barrier) lastprivate(sum)
{
  [...]
  #pragma omp barrier
  for (int j = 0; j < omp_get_num_threads(); ++j)
    sum += omp_get_value_from_sibling(partial_sum, j);
}
\end{ccode}
\end{samepage}
Each warp lane has its private copy of \cinline{partial_sum}, which are collecting partial sums, as in the preceding section but without limitation to a specific warp size.
Worksharing loops have an implicit barrier ensuring that all partial sums have been computed.
On Nvidia devices since Volta, it maps to the \cinline{__syncwarp} instruction or does nothing if the lanes are in lockstep.
The partial sums then are added up for the final result.
Here this is done by all lanes, but only the last task needs to do it because of \cinline{lastprivate}.
In CUDA, the library function \cinline{omp_get_value_from_sibling} can use the \cinline{__shfl_idx_sync} primitive.
A more efficient implementation could use \cinline{__shfl_down_sync}, either because the compiler recognizes the pattern or by exposing it as a function call.

Compared to the solution in~\cite{wang20-ompshuffle}, this has the downside that it has no fallback in case shuffle instructions are not supported; and implementing \cinline{omp_get_value_from_sibling} by other means would be inefficient.
A version that supports fallback using memory one level up would be the following, but it relies on compiler optimizations to promote \cinline{s[omp_get_thread_num()]} to a register.
\begin{ccode}
float sum = 0;
float s[omp_get_max_threads()] = {0};
#pragma omp parallel sync(shuffle,barrier) shared(s) lastprivate(sum)
{
  #pragma omp for
  for (int i = 0; i < n; ++i)
    s[omp_get_thread_num()] += A[i];
  sum = 0;
  #pragma omp barrier
  for (int j = 0; j < omp_get_num_threads(); ++j)
    sum += s[j];
}
\end{ccode}

\subsection{Level Partitioning}\label{sec:sublevels}

Previous examples assumed that a level occupies exactly one hardware level or collapses multiple levels into a virtual level.
We  propose the possibility to also split a hardware level.
Intel's GPU hardware natively supports changing the warp size~\cite{intel23-optguide}, but on Nvidia's GPUs it is fixed to 32 lanes.
However, we can divide a lane by two as shown below.
\begin{ccode}
#pragma omp parallel level(lanes) reserve(level(lanes(16)))
  #pragma omp parallel level(lanes(16))
  [...]
\end{ccode}
For the implementation this does not change a lot other than that it has to keep track of thread numbers and what nested directives bind to.
For instance, Nvidia's warp-level instructions provide arguments to bind only to a subset of warp lines but still execute on all lanes.
The \cinline{__shfl_*_sync} family of functions have \cinline{width} parameters to treat all warp subsections of that size as separate entities.

\subsection{Versioning}\label{sec:versioning}

Programmers can write multiple kernels optimized for different hardware.
OpenMP already offers the means to do this with the metadirective and \cinline{declare variant} directive.
For instance, a \cinline{device={isa("nvptx")}} filter can be used to  execute an optimized kernel for CUDA-based targets only where levels such as \dquote{warp} are available, and \cinline{device={arch("navi")}} selects a GPU generation from AMD.
Additionally, a robust application can fall back on a generic version of a kernel using the sync clause for hardware-independence.

\section{Language Extensions for Hierarchical Memory}\label{sec:mem}

In addition to instruction stream processors, hierarchy levels may have memory that is used transparently as a cache, be accessible faster from the processing unit than from sibling units (a NUMA domain), not accessible at all from sibling units (a scratch-pad), or each unit may use independent storage for the same virtual address\footnote{In the the OpenMP 6.0 draft called \cinline{groupprivate} memory, a generalization of \cinline{threadprivate}.}.

Using this nomenclature, OpenMP target device storage is a scratch pad (unless enabling unified shared memory), and the memory on it must be allocated by using either \cinline{omp_target_alloc} or the map clause.
This could motivate also supporting the map clause for lower-level memories such as block-shared memory.
Any access the mapped memory is redirected to the local memory, replacing the manual solution of declaring a temporary array in the outer construct and copying the data into it.

The more difficult problem is when each sibling can receive only part of the entire data because of memory constraints.
Also, when the data is written as well, how the data is written back at the end of the construct has to be defined.

The following illustration tiles a 2-dimensional array where each tile is distributed to a different device. 
\begin{ccode}
float A[1024][1024];
#pragma omp parallel level(devices:0-3)           \
    map(to(d):A[(d/2)*511:513][(d%2)*511:513])    \
    map(from(d):A[(d/2)*512:512][(d%2)*512:512])
\end{ccode}
The map clause is parameterized by a device number \cinline{d}, so each device can receive a different array section.
The sections  overlap so devices can access the immediate neighbors for reading, commonly referred to as a ghost surface.
A second map clause with a from-modifier specifies what elements to transfer back to parent memory that must have unique sources.
A reduction clause could be used to combine elements in case the array sections are overlapping.
To avoid wasting memory, the compiler will need to pack the transferred data into a new data layout and rewrite the address computations within the construct.
If the memory is a NUMA domain instead of a scratch-pad (so access from sibling devices are still possible, but slower), a similar syntax could be used to define data affinity.

Note that the target directive is not necessary here because \cinline{level(devices)} already declares our intention to offload.
Heterogeneous devices manifest as junctions in the hierarchy diagram, such as \cref{fig:h100}.
Without explicit device specification an implementation would likely execute this example on the host CPU.

For more complicated mapping scheduling, the only generic solution might be to implement the mapping manually by the programmer, for example, \cinline{A_dist[4][513][513]}, and then map each \cinline{A_dist[d]} to a different device.
In contrast to XcalableMP~\cite{lee12-xcalablemp}, which was explicitly designed for distributed memory (PGAS), our approach assumes that there is always a parent memory large enough to hold its child memories.

\section{Related Work}

The original hardware level names as invented by Nvidia were the grid, blocks, and threads.
Later programming models use different names to emphasize its vendor-independence: OpenMP calls them league/team/thread, OpenACC uses kernel/gang/worker, and OpenCL uses the names NDRange, work-group, and work-item.
In contrast to OpenMP which explicitly flattens the iteration space using the \cinline{collapse} clause, OpenCL's and CUDA's levels are multidimensional with up to three dimensions.
In the following we discuss how CUDA, SYCL, and OpenACC handle increasingly deeper hardware hierarchies.

\subsection{CUDA}\label{sec:rel_cuda}

Nvidia introduced \emph{Thread Block Clusters} in CUDA 11.8 with Compute Capability 9.0 as a new level between grid and block~\cite{nvidia-cuda90,nvidia17-cg}.
It is supported only with the Hopper generation and only with up to 8 blocks, which suggests it is using the common memory of a multiprocessor (see \cref{fig:h100}).
According to the documentation, however, they are required to be executed only on the same GPC.
Access to Thread Block Cluster memory is not a language extension but has to be done using the Cooperative Groups API.

Cooperative Groups~\cite{nvidia17-cg} is Nvidia's API to generalize the different abstraction levels inspired by \cinline{tiled_extent} from C++AMP.
It predefines six different levels, including multi-grid, which spans multiple devices, but also allows user-defined partitioning like we proposed in \cref{sec:sublevels}.
A Cooperative Group supports group-level operations such as barriers and shuffle methods (\cref{sec:shuffle}).

\subsection{SYCL}\label{sec:esimd}

SYCL 2020 added the notion of subgroups to the language~\cite{sycl2020}.
A subgroup usually represents the SIMD lanes of a warp but is specified only to be \squote{related} work-items.
Memory levels such as in \cref{sec:mem} are selectable for atomic using the \cinline{sycl::memory_scope} enumeration.
There is no specification of arbitrary depth, but vendors can and do introduce extensions that allow optimizing for their hardware, such as Xilinx \cinline{sycl::ext::xilinx::partition_ndarray}.

To support shuffle instructions, Intel proposed the ESIMD extension, which allows vector intrinsics instead of the implicit SIMT approach~\cite{intel-esimd2}.
That is, the user has to write the instructions that operate on subgroup lanes, usually using a C++ \cinline{simd} class---a vector with as many elements as the SIMD width; it also supports inline assembly.
Hierarchical partitions as shown in \cref{sec:sublevels} are supported using a \cinline{simd_view} class.
A noteworthy feature is to call ESIMD code from a SIMT context, by passing the code to the \cinline{invoke_simd} function. 
In the called function scalar parameters are pass as \cinline{simd} objects, unless declared as uniform.

\subsection{OpenACC}

OpenMP 3.3 introduced nesting multiple gang levels~\cite{openacc}.
The outer gang levels have to specify how many levels of gang parallelism are to be reserved for nested constructs.
This does not map to heterogeneous levels as in this proposal, but to the three dimensions of a CUDA grid.

\section{Conclusion}

OpenMP 1.0 started with a simple premise of flat, symmetric shared-memory parallelism.
As computing resources become more powerful, they are also becoming more diverse and complex, but the specification can react only after the new hardware emerges.
Devices must also match OpenMP's rigid execution model and hardware.
FPGAs do not, and hence an FPGA device will never be able to comply with the OpenMP specification, despite OpenMP priding itself on having been implemented for many different kinds of hardware~\cite{iwomp19-spectrum}.
With our proposal, there are fewer guarantees that an implementation must provide and greater flexibility for future hardware without releasing a new OpenMP specification.

Additionally, it has the advantage of being descriptive (just define the requirements of a level via the sync clause), similar to the loop construct without a loop, but also prescriptive when needed for performance optimization using a hardware-specific level in the level clause.
Programs that allocate compute hours on supercomputers generally require showing that the program has been optimized to the target hardware (e.g., \cite{doe23-incite,prace-cfp}).
With more complex hardware this will be increasingly harder to do if OpenMP does not provide the means to do so, and applications will have to use vendor-proprietary ecosystems (CUDA, HIP, DPC++, etc.) instead. 

Reusing the well-known \cinline{parallel} and \cinline{for}/\cinline{DO} constructs would be the most straightforward since we are reusing its principles of starting an SPMD region, then distributing work between them, but there may be too many legacy behavior conflicts may make introducing new directive names necessary.
It is also a major undertaking to revise OpenMP's current nomenclature of target, league, teams, contention groups, threads, and simd.
In any case, a major advantage over the current approach is the orthogonal relations between levels using properties, compared with the heterogeneous but fixed levels and their pairwise defined relations that were intended to match a now-outdated generation of hardware.

\subsubsection{Acknowledgments}
The idea of reusing the parallel directive instead of teams was first brought up by Bronis de Supinski.

This research was supported by the Exascale Computing Project (17-SC-20-SC), a collaborative effort of the U.S. Department of Energy Office of Science and the National Nuclear Security Administration, in particular its subproject SOLLVE.


\appendix

\section{Nvidia Grace Hopper Superchip Hierarchy}\label{sec:hardware}\label{sec:h100}

In this section, we illustrate the complexity and hierarchical depth of contemporary high-performance hardware using Nvidia's most recent GPGPU design, the Hopper architecture~\cite{nvidia-h100}, as an example.
GPGPU hardware from other vendors such as Intel and AMD generally expose a similar hierarchy and implement the same programming model assumed by OpenMP, SYCL, OpenACC, etc., but will also be different enough to require different optimization strategies.
For instance, Nvidia hardware has 32 lanes per warp, while AMD has 64, and Intel's GPU accelerators are configurable between 1 and 32 lanes per warp.
Other typical differences are the number of levels, the amount of memory per level, and the number of tasks supported on each level.
In contrast, specialized hardware such as FPGAs and dedicated AI accelerators may diverge significantly.

The current OpenMP programming model does not account for these differences and uses a one-model-to-match-them-all approach which cannot accurately represent the diversity of accelerator architectures.
Moreover, compute hardware can be expected to become more complex in unpredictable ways in the future.

\tikzset{omptarget/.style={draw=jonquil!80,line width=1.2pt,rounded corners,top color=jonquil!40,bottom color=jonquil!20,shading angle=55}}
\tikzset{ompteamsdistribute/.style={draw=red,line width=1.2pt,rounded corners,top color=red!30,bottom color=red!10,shading angle=55}}
\tikzset{ompparallelfor/.style={draw=blue,line width=1.2pt,rounded corners,top color=babyblueeyes,bottom color=babyblueeyes!50,shading angle=35}}
\tikzset{ompsimd/.style={draw=green,line width=1.2pt,rounded corners,top color=green!40,bottom color=green!20,shading angle=25}}
\tikzset{sequential/.style={draw=gray,line width=1.2pt,rounded corners,top color=gray!40,bottom color=gray!20,shading angle=15}}

\tikzset{memory/.style={database,database radius=3mm,database segment height=1.5mm,database top segment={top color=blue!10,bottom color=white,shading angle=15},database middle segment={fill=white},database bottom segment={fill=white},outer xsep=2mm}}

\begin{figure}[tb!]
\begin{resizepar}
\begin{tikzpicture}
\tikzset{level/.style={align=center,node distance=3mm}}
\tikzset{consistsof/.style={->}}

\node[level](node) {\textbf{Cluster Node}\\8 Superchips};
\node[level,below=of node](superchip) {\textbf{Nvidia Grace Hopper}\\1 CPU, 1 GPU};

\matrix[ampersand replacement=\&,nodes={level},column sep=5mm,below=2mm of superchip] {
  \node[](cpu) {\textbf{Nvidia Grace CPU}\\72 Cores\\117MB (L3) Cache, 512GB DDR Memory};         
  \node[below=of cpu](core) {\textbf{Arm Neoverse V2}\\4 SIMD Units\\64KB (L1) + 1MB (L2) Cache};
  \node[below=of core](simdunit){\textbf{SVE Instruction}\\512 Bit Registers};      
  \node[below=of simdunit] (simdfu){\textbf{SIMD Unit}\\128 Bit Subregisters};
  \node[below=of simdfu] (lane){\textbf{SVE Lane}};
\&
  \node[](gpu) {\textbf{Nvidia H100 GPU}\\schedules Kernels\\96GB HBM3 Memory};                                      
  \node[below=of gpu](kernel) {\textbf{Compute Kernel (\dquote{CUDA Grid})}\\2 Partitions}; 
  \node[below=of kernel](partition) {\textbf{Cache Partition}\\4 GPCs\\30MB Cache (L2)};                      
  \node[below=of partition](gpc) {\textbf{GPU Processing Cluster (GPC)}\\9 TPCs}; 
  \node[below=of gpc](tpc) {\textbf{Texture Processing Unit (TPC)}\\2 SMs}; 
  \node[below=of tpc](sm){\textbf{Streaming Multiprocessors (SM)}\\4 Warp Schedulers\\256KB combined Shared/(L1) Cache};
  \node[below=of sm](dispatch) {\textbf{Warp Scheduler + Dispatch Unit}\\schedules Warps\\64KB Register File};         
  \node[below=of dispatch](cta) {\textbf{Cooperative Thread Array (CTA; \dquote{CUDA Block})}};
  \node[below=of cta](warp) {\textbf{Warp}\\32 Lanes\\32 FP32 + 16 FP64 + 1 Tensor Cores};
  \node[below=of warp](gpufu) {\textbf{FP32/FP64/Tensor CUDA Core}};
  \node[below=of gpufu](thread) {\textbf{Warp Lane (\dquote{CUDA Thread})}\\Floating-Point or 32 Bit Integer Registers};
  \node[below=of thread](vlane) {\textbf{Integer SIMD Intrinsic Lane}};
    \\
};

\path (node) edge[consistsof] (superchip);

\path (superchip) edge[consistsof] (cpu);
\path (cpu) edge[consistsof] (core);
\path (core) edge[consistsof] (simdunit);
\path (simdunit) edge[consistsof] (simdfu);
\path (simdfu) edge[consistsof] (lane);

\path (superchip) edge[consistsof] (gpu);
\path (gpu) edge[consistsof] (kernel);
\path (kernel) edge[consistsof] (partition);
\path (partition) edge[consistsof] (gpc);
\path (gpc) edge[consistsof] (tpc);
\path (tpc) edge[consistsof] (sm);
\path (sm) edge[consistsof] (dispatch);
\path (dispatch) edge[consistsof] (cta);
\path (cta) edge[consistsof] (warp);
\path (warp) edge[consistsof] (gpufu);
\path (gpufu) edge[consistsof] (thread);
\path (thread) edge[consistsof] (vlane);

\begin{pgfonlayer}{backbackbackground}
\node[tight,fit={(node) (superchip) (cpu)},ompparallelfor] {}; 
\node[tight,fit={(core) (simdunit) (simdfu)},ompsimd] {}; 
\node[tight,fit={(lane)},sequential] {};
\end{pgfonlayer}

\begin{pgfonlayer}{background}
\node[tight,fit={(kernel) (partition) (gpc) (tpc) (sm) (dispatch)},ompteamsdistribute](grid) {}; 
\node[tight,fit={(cta) (warp) (gpufu)},ompparallelfor](block) {}; 
\node[tight,fit={(vlane)},sequential] {};
\end{pgfonlayer}
\begin{pgfonlayer}{backbackground}
\node[fit={(gpu) (kernel) (grid) (block) (thread) (gpufu) (vlane)},omptarget] {}; 
\end{pgfonlayer}

\matrix[matrix of nodes,ampersand replacement=\&,anchor=south west,nodes={anchor=west},label={above:Typical OpenMP Mapping}] at (current bounding box.south west) {
 |[omptarget]| \phantom{XX} \& \texttt{\#pragma omp target} \\
 |[ompteamsdistribute]| \phantom{XX} \& \texttt{\#pragma omp teams distribute} \\
 |[ompparallelfor]| \phantom{XX} \& \texttt{\#pragma omp parallel for} \\
 |[ompsimd]| \phantom{XX} \& \texttt{\#pragma omp simd} \\
 |[sequential]| \phantom{XX} \& Sequential code \\
};

\end{tikzpicture}
\end{resizepar}
\caption{Hierarchical parallelism of a fictitious Nvidia H100 cluster.}
\label{fig:h100}
\vspace*{-6mm}
\end{figure}
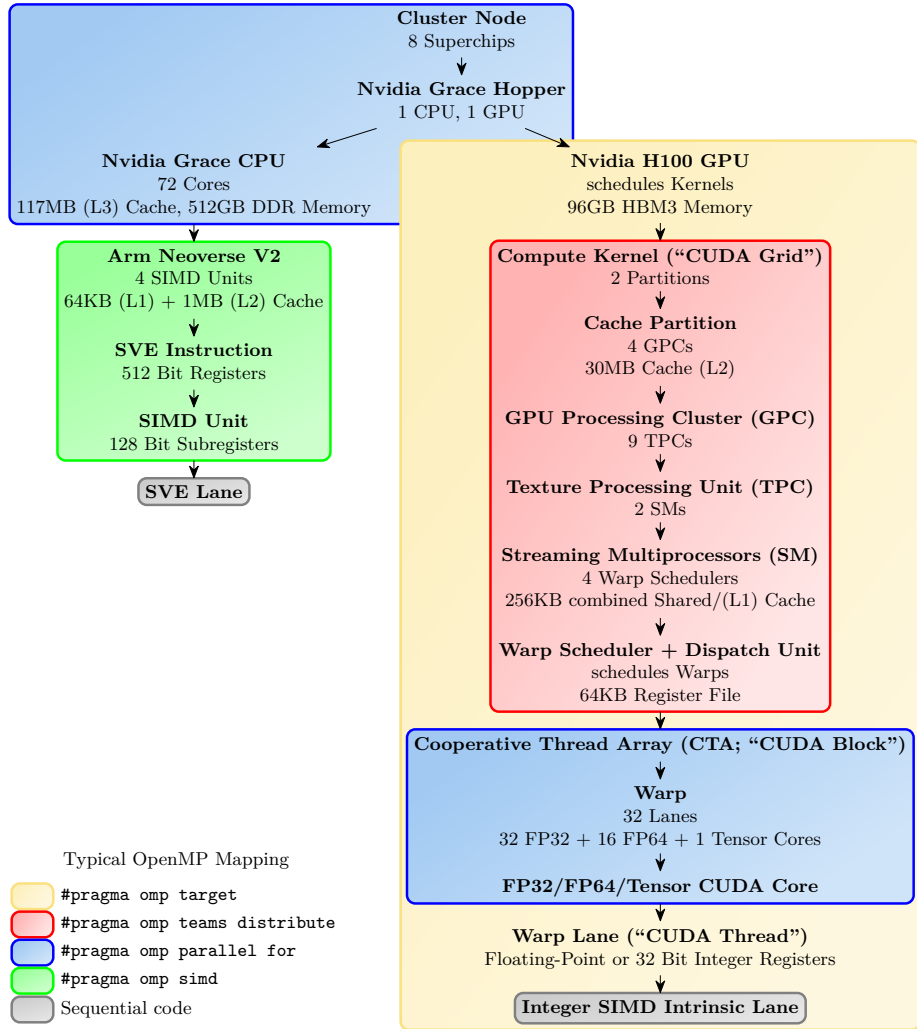

\Cref{fig:h100} shows the H100 in a hypothetical configuration that might be sold as the Nvidia HGX Grace Hopper system or be part of the LANL Venado cluster.
Unlike previous chips, Nvidia also offers its H100 \squote{Hopper} architecture chips in combination with an ARM-based CPU called \squote{Grace} integrated into the elegantly named \squote{Grace Hopper} superchip~\cite{nvidia-gracehopper}.
%
Nvidia points out that the Grace CPU does not have NUMA for \dquote{high developer productivity}, but in a typical setting where multiple processors are connected via NVLink, including Nvidia's own HGX offerings, each superchip is effectively its own NUMA domain.
Otherwise, the Grace CPU matches the architectures that are taken account with the \cinline{parallel for} construct and \cinline{proc_bind} clause and the \cinline{simd} construct.

In contrast, the Hopper GPU has many more levels than the \cinline{teams distribute} and \cinline{parallel for} constructs can represent.
An H100 chip is not just a flat collection of multiple SMs but builds a hierarchy itself.
Most of them may not be relevant for either the programming model or performance, making them transparent to the programmer, except for the L2 cache, which splits the GPU into halfs.

A H100 multiprocessor (SM) is a multicore processor itself and can process 4 instruction streams in parallel, each with its own warp scheduler and register file.
The instruction streams can come from either the same or different Cooperative Thread Arrays (CTAs) if the memory constraints allow, but a currently resident task will never move to another SM.
Since the four schedulers share the same local memory, data transfers between the warps on the same SM is significantly faster than between warps on different SMs.
An algorithm may be able to take this into account; but since the same memory is also used for a per-block shared memory, it is architecturally possible to directly access another block's scratch pad memory when executing on the same SM using the abstraction called \squote{\emph{distributed} shared memory}.
In OpenMP, it is currently not possible to access this memory.

A Cooperative Thread Array itself is split into warps of 32 lanes each, but whether they are executed in parallel depends on the hardware.
In contrast to AMD and Intel, handling of divergent lanes of the same warp is done by the hardware (instead of instructions that explicitly set the execution mask) and, beginning with the Volta architecture, features \squote{independent thread scheduling}~\cite{nvidia-v100}.
That is, the hardware is allowed to decide itself when to mask out a lane and when to continue its execution.
As a result, which lanes are executed in lockstep is not predictable in software anymore, and Nvidia added \cinline{__syncwarps} and additional \cinline{_sync}-suffix instructions to CUDA to give software some control over the processes.

With 32 FP32 cores, the H100 can execute all warp lanes at once in the case of a single-precision instruction, but since only 16 FP64 cores are available, it takes two rounds for all 32 lanes of a double-precision instruction to execute.
Neither CUDA nor OpenMP exposes this detail, but some algorithms might be able to exploit this detail knowing that using two warps each using just 16 lanes may result in the same performance as a full warp or that lanes that are not in the same group of 16 can diverge without performance penalty.
Incidentally, this similarly applies to the Grace CPU, which processes a single 512-bit SVE instruction independently in 4 SIMD units.

In addition to the SIMT programming model, CUDA supports a limited selection of traditional SIMD intrinsics, but only for integer operations~\cite{nvidia-cuda90}.
For instance, the \cinline{__vadd2} intrinsic treats a 32-bit integer register as a vector register containing two 16-bit integers and adds them independently.
Similarly, \cinline{__vadd4} adds four 8-bit integers.
We are not aware of any compiler that emits these vector instructions without explicitly using the intrinsics, even when using \cinline{#pragma omp simd}.
\bibliographystyle{splncs04}
\bibliography{bibliography}

\end{document}